\documentclass[11pt,twoside]{article}
\usepackage{asp2010}

\resetcounters

\begin{document}

\title{Observations that can unravel the coherent radio emission mechanism in pulsars}
\author{Dipanjan Mitra}
\affil{National Centre for Radio Astrophysics, TIFR, Pune University 
Campus, Post Bag 3, Ganeshkhind Pune 411007, INDIA }

\begin{abstract}
Searching for the physical mechanism that can excite the coherent radio emission in 
pulsars is still an enigmatic problem. A wealth of
high quality observations exist, which over the years have been instrumental in
putting stringent constraints to pulsar emission models. In this article we will 
discuss the observational results that strongly suggests that pulsar radio emission is
excited by coherent curvature radiation. We will also mention issues that remain
to be resolved.
\end{abstract}

\section{Introduction}
Magnetospheric coherent radio emission from pulsars consist of a main pulse --
which is the most bright structure in the pulse profile and associated with 
a linear polarization position angle (PPA) swing across the pulse; 
sometimes an interpulse -- which is located 180$^{\circ}$ away from the main 
pulse and also associated with a PPA swing; occasionally pre/post--cursor emission 
which is a highly polarized temporal structure with a flat PPA connected via a 
bridge to the main pulse but located significantly away from the main pulse
and the recently discovered off--pulse emission (Basu etal. 2011) which is like a broad 
emission component observed in regions where no obvious temporal structures
are seen in a pulse profile. 

To date there is no self--consistent theory that can explain the overall 
aspect of pulsar emission. Most theories use that idea that the 
region around the neutron star is a charge--separated magnetosphere
which is ``force free'', meaning that the electromagnetic energy 
is significantly larger than all other inertial, pressure 
and dissipative forces. The magnetosphere is initially charge 
starved and supply of charged particles can come from the neutron 
star or due to pair creation in strong magnetic fields.
Subsequently the magnetosphere attains charge neutrality by accelerating
particles with density equal to the Goldreich--Julian density.  
In all models of pulsar radio emission there is general agreement that 
the radio emission arises due to growth of plasma 
instabilities in the relativistic plasma streaming along 
curved magnetic field lines. Such processes in pulsar astrophysics are: 
cyclotron maser (Kazbegi, Machabeli \& Melikidze 1991), 
two--stream instabilities (Usov 1987; Asseo \& Melikidze 1998), 
collapsing solitons (Weatherall 1998), charged
relativistic solitons (Melikidze, Gil \& Pataraya 2000, hereafter MGP00; Gil, 
Lyubarsky \& Melikidze 2004, hereafter GLM04) and linear acceleration maser (Melrose 1978).

The aim of this article is to summarize the key observational results
that have given or have the possibility of providing constraint
on understanding the coherent radio emission from pulsars. 
Here we will concentrate on the main pulse emission
and will also take the position that the coherent radio emission mechanism is 
excited by curvature radiation from charged bunches (like solitons).
The physical process include formation of a inner vacuum gap (IVG) near the 
pulsar polar cap where non-stationary spark associated 
relativistic ($\gamma_p \sim 10^6$) primary particles are generated 
(Ruderman \& Sutherland 1975, hereafter RS75).
These particles further radiate in strong magnetic field and the photons thereby 
produce secondary e$^+$e$^-$ plasma with $\gamma_s \sim 400$. The 
two--stream instability in the plasma 
generates Langmuir plasma waves and the modulational 
instability of the Langmuir waves 
leads to the formation of charged solitions which can excite
extraordinary (X) and ordinary (O) modes of curvature radiation in the plasma 
as shown by  MGP00 and GLM04. For this process to work, highly non-dipolar
surface magnetic field is essential (Gil, Melikidze \& Mitra 2002).

\section{The Shape of the main pulse emission region}
The radio emission from pulsars almost invariably arises from 
regions of open dipolar field lines. The linear PPA swings for a large
number of pulsars are in very good agreement with the rotating--vector
model (RVM, Radhakrishnan \& Cooke 1969) which predicts the behaviour of the 
PPA arising from open dipolar field lines. The average PPA in several pulsars
have a complex non-RVM behaviour, however they are mostly complicated
due to the presence of orthogonal polarization modes (OPM). Single pulse
polarization can be used to separate the OPMs after which the RVM is clearly 
satisfied for the individual modes (Gil \& Lyne 1994).
The distribution of pulse width with rotation period shows a lower bound 
which scales with pulsar period as $P^{-0.5}$, which
reflects the change of dipolar open field lines due to the changing 
light cylinder distance. Most pulse profiles have one or many subpulses or components.
Detailed phenomenological study of average pulse shape and polarization 
reveal that the pulsar emission beam has a central core emission (surrounding 
the magnetic axis, Rankin 1990) with 
typically two or three nested cones around them (Rankin 1993, Mitra \& Deshpande 1999). 
Each of the cones scales as $P^{-0.5}$ and the observed 
subpulses are the line of sight cuts across the core or conal regions. 
The distribution of the subpulse width of the 
core or cone follow a lower bound 2.4$^{\deg} P^{-0.5}$ (Maciesiak etal. 2012), 
where 2.4$^{\deg}$ equals the polar cap size for a 1 second period pulsar 
and a neutron star with a radius of 10 km. In this core-cone model, it is important to 
understand that the nested cones are not uniformly illuminated, and hence
a given intensity pattern in a pulsar can have significant variation in component
intensity, however the location of the components appear to follow the 
conal structure. 

The cone structure in the RS75 model results due to the $E \times B$ drift of plasma 
columns or ``sparks" produced in the IVG. The core emission
is the central spark (Gil \& Sendyk 2000) while the other 
sets of sparks populate themselves in the polar cap maintaining a 
distance between them which is of the order
of the vacuum gap height. The sparks themselves have the VG height dimension.
While theoretically the formation of the IVG 
is possible, the physical mechanism of how exactly 
the sparks populate the polar cap and develop in size is not clear.
Nonetheless, if these sparks eventually generates a streaming flow of plasma and
can generate coherent radio emission at about 500 km from the neutron 
star surface, then it is possible to explain the observed subpulse widths and
core-cone structure observed in pulsar.

\articlefiguretwo{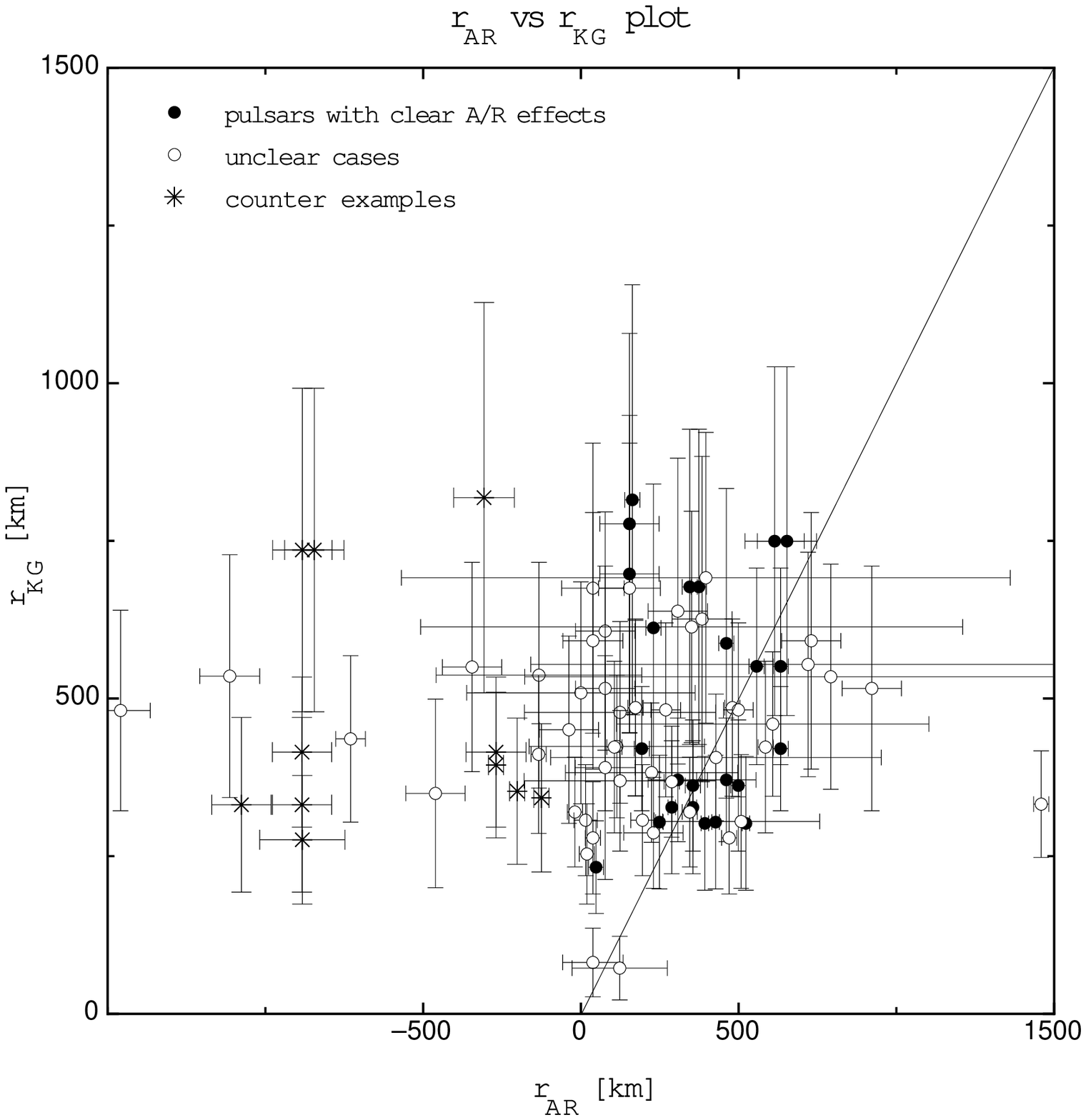}{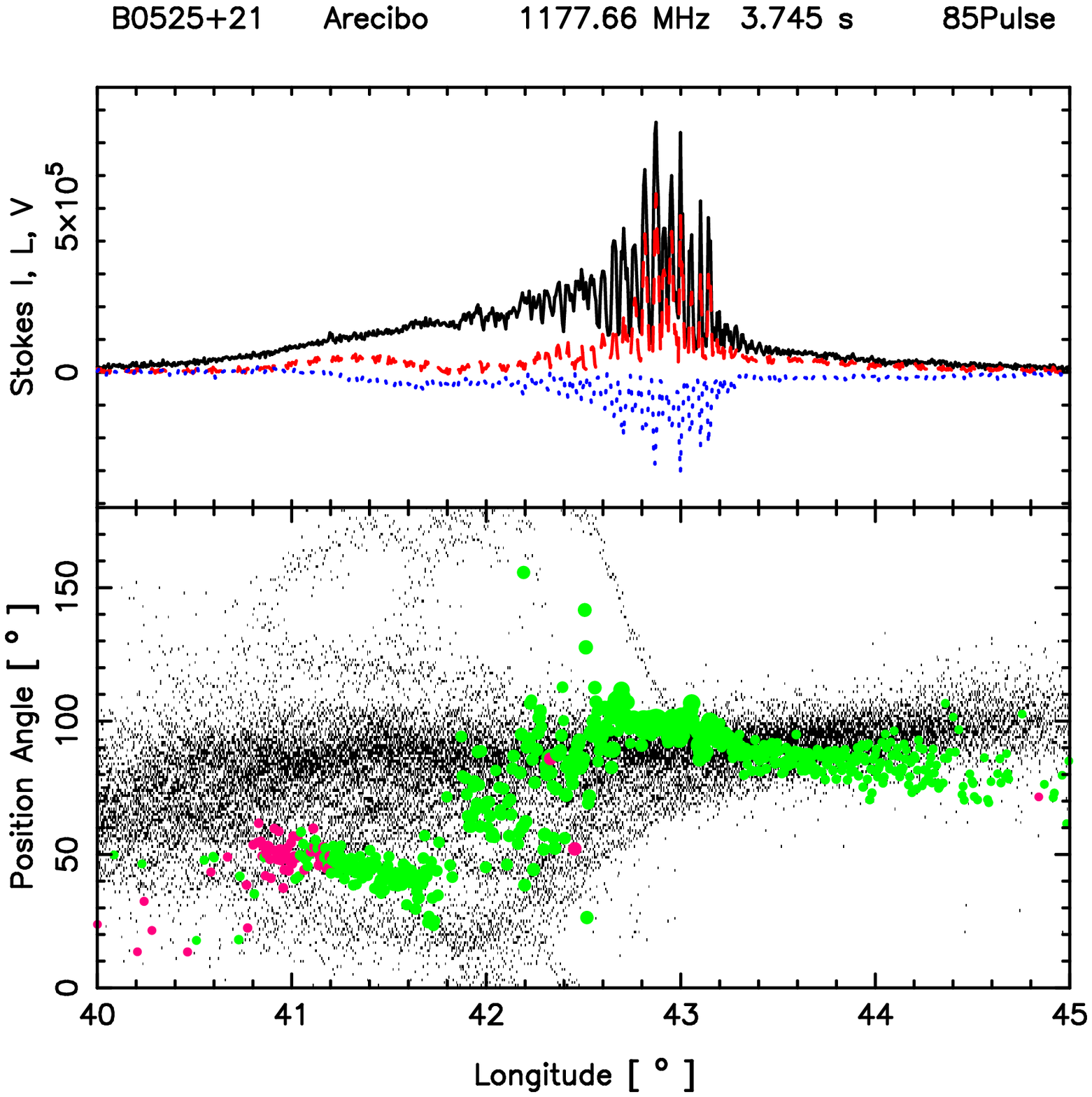}{height}{The left figure 
(adopted from Krzeszowski etal. 2009) 
show the comparison between radio emission heights estimated
from the geometrical (yaxis) and delay method (xaxis). The conal emission in 
pulsars is consistent with emission arising at around 500 km 
(refer Krzeszowski etal. (2009) for details). The right panel shows 
an Arecibo observation of a single pulse for the 3.7--sec pulsar 
PSR B0525+21, where microstructure of around 180 $\mu$sec is seen 
(Backus, Mitra \& Rankin 2012, in preparation): a much smaller 
timescale than the angular beaming timescale of 3 millisec.}

\section{The emission height of the main pulse}
Two methods of approach have provided emission--height 
determinations in pulsars, namely the geometrical method and 
the delay method. In the geometrical method the PPA traverse is used 
to infer the magnetic axis inclination angle and the line of 
sight angle, and using the pulse width dimension along 
with model of the open dipolar field lines, the height can be estimated.
The delay method as was suggested by Blaskiewicz etal. (1991) 
is based on the kinematical effect of
aberration and retardation (A/R) and subsequent careful 
derivation shows that emission heights can be estimated 
independent of pulsar geometry (see Dyks etal. 2004). 
The A/R effect is seen as a shift between the center of the total intensity
profile and the fiducial plane containing the magnetic and spin axis
which is often identified as the steepest gradient point of the PPA traverse 
or the peak of the core emission. The merits/demerits and usage of the 
height estimation methods can be found in Mitra \& Li (2004) and 
Dyks etal. (2004). One important point to note here is that 
the methods mentioned here can only give emission heights for the 
conal emission region. Determination of core emission height
has not been possible to date.

A few notable works dedicated to finding emission heights using the geometrical
method are: Rankin (1993), Kijak \& Gil (1998) 
and the delay heights:  BCW, 
von Hoenchbroech et al (1999), 
Malov \& Suleimanova (2000), Gangadhara \& Gupta (2001), Krzeszowski etal. (2009).
The left panel of Fig~\ref{height} shows a comparison between the 
geometrical and delay method by Krzeszowski etal. (2009), where one 
can clearly see that the emission arises from about 500 km 
above the neutron star's surface. This finding is a very significant input 
to the pulsar emission--mechanism problem.
The only plasma instability that can grow at these heights (where the magnetic 
field is very strong and the plasma is constrained to move along the 
magnetic field) is the two--stream instability. Hence models like 
cyclotron maser, which can only give rise to the coherent radio emission
near the light cylinder, can be ruled out.

\section{Single pulse dynamics of the main pulse}
A range of phenomena occuring at different time scales are observed in pulsars. 
The ones which are intrinsic to the pulsar emission are nulling, moding, 
drifting and microstructure. In pulsar nulling the radio emission
suddenly switches off for time scales as short as a pulsar period 
up to a few weeks to months. During pulsar moding 
the average pulse profile suddenly changes from one stable form to another
and a given mode can last for intervals of a few minuites to hours. The phenomena
of nulling and moding are perhaps the most difficult emission phenomenon
to explain and any further discussion on this is beyond the scope of this article. 

Pulsar drifting and microstructure phenomena
gives indirect hints about the radio loud plasma.    
Drifting subpulses exhibit drif through the 
average profile in a very regular manner. In fact for 
several pulsars very detailed analysis of the observations 
reveal that a given subpulse seem to have a periodic behaviour
which can be modeled as a circular carousel 
of emitting sparks (e.g. Deshpande and Rankin 2001; Mitra and Rankin 2008),
while there are quite a few other pulsars where the carousel
timescale can be inferred. In the RS75 model the carousel rotation is explained as 
$E \times B$ drift of the spark--associated plasma columns in the IVG. 
However, it is found that the observed timescale of carousel 
rotation are much longer (several tens of seconds) than the RS75 prediction.
A major refinement of this model was given by Gil, Melikidze \& Geppert (2006), 
where the longer carousel rotation time was explained by a partially screened 
vacuum gap. They also estimated the thermal 
X--ray emission that arises due to bombardment of charged particles
created in the VG onto the neutron star surface. Such thermal X--ray emission has been 
found in several isolated neutron stars and its connection to carousel rotation 
is an active area of pulsar research. Additionally, these observations provide
direct evidence to the model that an IVGs populated with sparks
exist in pulsars. 

The pulsar microstructure phenomenon is observed as short time--scale features 
in the single pulse ranging from 1 to several hundreds of microsec 
(Cordes 1979; Lange etal. 1998). 
The microstructure has been thought to reveal either the Lorentz factor
of the emitting plasma (Cordes 1979) or are signatures of plasma disturbances 
(Weatherall 1998).
Cordes (1979) finds that the average microsturcture timescale ($t_{\mu}$) 
scales with pulsar period as $t_{\mu} \sim 10^{-3} P$. If we assume that
pulsar microstructures result due to the angular beaming of relativistic 
particles, then we obtain a Lorentz factor of around 150 based on the 
Cordes (1979) relation. We have however ourselves tried to establish 
this relation but have not been successful so far. We also find that 
even for very long period pulsars (see Fig~\ref{height}), in some bright single pulses
the microstructure scale is much smaller than the angular beaming time scale.  
The theory to understand these short timescale temporal 
effects is still not fully developed. 
However, microstructures in pulsars are the best examples
of the smallest spacial and temporal scale plasma variations producing coherent
radio emission. In the IVG model, the microstructures corresponds
to spark--associated plasma columns of secondary plasma
with Lorentz factors of about 100--500. A large number of such sparks 
add up incoherently to produce a given pulsar subpulse.
 
\section{Orientation of the escaping waves of the main pulse}

Lai etal. (2001) used the x-ray image of the Vela pulsar wind nebula 
and absolute PPA to establish that the 
electric vector emanating out of the pulsar is orthogonal to the magnetic field
planes, and hence represents the extraordinary (X) mode.  
This significant observational result for the
first time demonstrated that the electric fields emerging
from the Vela pulsar magnetosphere are perpendicular to the dipolar
magnetic field planes.  Lai etal. (2001) also showed that
the proper motion direction (PM) of the pulsar is aligned with the
rotation axis. Johnston etal. (2005) \& Rankin (2007) produced a distribution of $\mid$PM- absolute PPA$\mid$
for a few pulsars and found a bimodal distribution around zero and 90$^{\circ}$. 
Assuming that the pulsars PMs are parallel to the rotation axis, the bimodality
could be explained as emerging radiation being either parallel or perpendicular
to the magnetic field planes, since pulsars are known to have orthogonal 
polarization modes.  
Alternatively PMs of pulsars can also be parallel or perpendicular to the rotation axis.
While both the above explanations are possible, it is clear that the electric vectors
of the waves which detache from the pulsar magnetosphere to reach the observer 
follows the magnetic field planes. 
This is in agreement with the IVG class of models where
MGP00 and GLM04 demonstrate that curvature radiation can excite the X and O modes
in plasma at around 500 km, and the X mode can escape the pulsar magnetosphere almost as in vacuum 
and reach the observer. 

\section{Pulsar Polarization and adiabatic walking condition (AWC)}
Mitra, Gil \& Melikidze (2009) argued that single pulses with 
close to 100\% linear polarization are most suitable
for unraveling the pulsar emission mechanism.
They showcased highly polarized single pulses from several pulsars where the 
PPA followed the rotating--vector model.
These pulses, which are relatively free from depolarization, must consist exclusively 
of a single polarization mode which they associate with the X-mode
excited by the coherent--curvature radiation. 
This argument however only holds good if the  wave polarization
at the generation point in the magnetosphere does not modify as it propagates
in the magnetosphere. Cheng \& Ruderman (1979) argued that if the 
AWC (which is given by $|\Delta N|kl>>1$, where $\Delta N$ 
is the difference in index of refraction $N=ck/\omega$ between the X and O modes) 
is satisfied, then the wave polarization slowly rotates, and hence as
the wave detaches from the magnetosphere, it no longer carries information 
about the generation point.  However, based on a rigorous treatment of 
the radiation mechanism GML04 and Melikidze, Mitra \& Gil (2012 in preparation) 
argue that the AWC does not hold in the pulsar magnetosphere and hence the 
X--mode can escape the pulsar magnetosphere unaffected. 

There are two aspects of pulsar polarization which are difficult to understand. One 
is the existence of OPMs where both the X and O modes are present. 
Most theories predict that the O-mode should be damped in the
magnetosphere. How it escapes is still a puzzle to be solved.
The second issue regards circular polarization. If the emission 
results from incoherent addition of smaller coherently emitting 
units, then one does not expect any
phase relation between parallel and perpendicular electric fields
and hence the circular polarization should vanish. Often propagation 
effects are invoked to explain circular polarization, 
however these explanations ignore a wide range of other phenomenon, and we feel
that currently there is no good explanation for the circular polarization 
behaviour in pulsars.

\section{Summary}
The formation of IVG leading to radio emission excited by 
coherent-curvature radiation is by far the most successful 
theory that explains the main pulse emission phenomena.
Currently challenging simultaneous X-ray and radio observations are being 
done to understand the IVG conditions in pulsars. We are still 
uncertain about the origin of pre/post-cursor and off-pulse emission
in pulsars. More observations are needed to search and characterize such emission.
It is possible that different coherent radio emission mechanisms are responsible
for such emission. 

\acknowledgements I would like to thank my esteemed colleagues 
- Joanna Rankin, Janusz Gil \& G. Melikidze, with whom I have worked extensively 
on the pulsar emission problem over the years.

\end{document}